\documentclass[prd,
,superscriptaddress,
nofootinbib,%
tightenlines ]{revtex4}

\usepackage{epsfig}
\usepackage[colorlinks=true,linkcolor=blue,urlcolor=blue,citecolor=blue]{hyperref}
\usepackage{amsmath}
\usepackage{amsfonts}
\usepackage{amssymb}
\usepackage{slashed}
\usepackage{mathtools}
\newcommand{\beq}{\begin{equation}}
\newcommand{\eeq}{\end{equation}}
\newcommand{\beqa}{\begin{eqnarray}}
\newcommand{\eeqa}{\end{eqnarray}}

\newcommand{\ben}{\begin{displaymath}}
\newcommand{\een}{\end{displaymath}}
\newcommand{\be}{\begin{equation}}
\newcommand{\ee}{\end{equation}}
\newcommand{\bea}{\begin{eqnarray}}
\newcommand{\eea}{\end{eqnarray}}

\usepackage{epsfig}
\begin{document}
\preprint{MKPH-T-04-14}
\title{Gravitational form factors of the deuteron}

 \author{J.~Yu.~Panteleeva}
  \affiliation{Institut f\"ur Theoretische Physik II, Ruhr-Universit\"at Bochum,  D-44780 Bochum,
 Germany}
\author{E.~Epelbaum}
 \affiliation{Institut f\"ur Theoretische Physik II, Ruhr-Universit\"at Bochum,  D-44780 Bochum,
 Germany}
 \author{A.~M.~Gasparyan}
 \affiliation{Institut f\"ur Theoretische Physik II, Ruhr-Universit\"at Bochum,  D-44780 Bochum,
 Germany}
\author{J.~Gegelia}
 \affiliation{Institut f\"ur Theoretische Physik II, Ruhr-Universit\"at Bochum,  D-44780 Bochum,
 Germany}
 \affiliation{Tbilisi State
University, 0186 Tbilisi, Georgia}

 \date{25 November, 2024}
\begin{abstract}

The gravitational form factors of the deuteron are calculated in the
framework of non-relativistic chiral effective field theory.  
Non-relativistic reduction of the matrix element of the
energy-momentum tensor operator for spin-one systems is worked out,
and the gravitational form factors of the deuteron are extracted from
the three-point function of the energy-momentum tensor
using the LSZ reduction formula.
The obtained form factors are compared to results of model
calculations available in the literature.

\end{abstract}

\maketitle

\bigskip
\bigskip
\begin{center}

{\large This paper is dedicated to memory of our dear teacher, friend and colleague Maxim Polyakov. \\
  It provides new insights into the structure of hadronic and nuclear
  systems encoded in the gravitational form factors, the activity
  that has been initiated by Maxim.} 
\end{center}

\bigskip
\bigskip

\section{\label{introduction}Introduction}

It's a great honor to contribute to this special issue of Acta Physica
Polonica B dedicated to the memory of Mitya Diakonov, Vitya Petrov and
Maxim Polyakov. These virtuosos of theoretical physics left deep and
lasting footprints in the field of QCD, and they also played a
central role in shaping the research directions of our institute.
One of us (E.E.) came across Mitya, Vitya and Maxim in the middle of
1990s, when he came to Bochum as a student. While working on
different topics, it was a truly unforgettable experience to enjoy a
unique, scientifically-vivid atmothphere of the TPII-institute
with lively discussions in the ``Strong interaction room''
and intense Russian-style seminars. For a personal recollection of
this time see also a recent paper by another former TPII member Hyun-Chul Kim
\cite{Kim:2024tae}.  

In the last decade, the main scientific connection between Maxim's and our
groups at the TPII-institute was, of course, the chiral physics. Over
the years, Maxim was also continuously interested  (among a plethora of
other things) in probing internal structure of hadrons with
gravity-induced interactions, see
\cite{Polyakov:2002yz,Polyakov:2018zvc} for some of his seminal
contributions to the 
field. Naturally, he came to the idea to approach this problem using the
language of the effective chiral Lagrangian, generalized to curved
space-time \cite{Alharazin:2020yjv}. This was how our way-too-short
collaboration with Maxim on this topic started \cite{Gegelia:2021wnj, Epelbaum:2021ahi,Epelbaum:2022fjc}. 

This paper represents a next logical step towards uncovering the structure
of strongly interacting systems with gravitational
probes. Specifically, we focus here on the gravitational form factors
(GFFs) of the simplest nuclear system, the deuteron, using 
the framework of 
chiral effective field theory (EFT). Triggered by the seminal papers 
by Weinberg \cite{Weinberg:rz,Weinberg:um}, it offers a systematic
formalism with a controlled improvable accuracy. For recent
reviews see, e.g., Refs.~\cite{Epelbaum:2008ga,Machleidt:2011zz,Hammer:2019poc,Epelbaum:2019kcf}.
While the electromagnetic structure of the deuteron has already been extensively
analyzed in the EFT framework using various approaches
\cite{Kaplan:1998sz,Phillips:1999am,Walzl:2001vb,Phillips:2003jz,Phillips:2006im,Kolling:2012cs,Epelbaum:2013naa,Filin:2019eoe,Filin:2020tcs},
the GFFs have, to the best of our knowledge, been only considered in
model calculations using
a phenomenological nucleon-nucleon potential~\cite{He:2023ogg,He:2024vzz}, see also Ref. \cite{Freese:2022yur} for a related discussion.
In the present work we fill this gap and extract the GFFs of the
deuteron by calculating the three-point function of the
energy-momentum tensor (EMT) in chiral EFT and applying the
Lehmann-Symanzik-Zimmermann (LSZ) reduction formalism in analogy to
Refs.~\cite{Kaplan:1998sz,Epelbaum:2013naa}. We employ the
non-relativistic formulation of chiral EFT with pions and nucleons as
the only dynamical degrees of freedom.

Our calculation is restricted to the leading-order (LO) nucleon-nucleon (NN) potential
which, according to Weinberg's power counting, is given by 
derivative-less contact interactions and the one-pion-exchange. 
Non-relativistic expressions for the EMT insertions at LO,
next-to-leading order (NLO) and  next-to-next-to-leading order (NNLO)
are obtained from the corresponding Lorentz-invariant expressions by
applying the standard heavy baryon reduction.  


Our paper is organized as follows. In section \ref{sec2}
we briefly outline a general  formalism to calculate the deuteron form
factors in quantum field theory. The integral equations for the deuteron structure functions are
discussed in section \ref{sec3}, while the actual calculation of the
GFFs is presented in section \ref{calc}.
A brief summary of the most important results of this study is
provided in section \ref{sec4}.

\section{Gravitational form factors of the deuteron}
\label{sec2}

Matrix elements of conserved EMT operator for spin-1 systems can be
parameterized in terms of six GFFs -- the coefficient functions
$A_{0, \, 1}(q^2) $, $D_{0,\, 1}(q^2) $, $J(q^2) $ and $E(q^2)$ of
the independent conserved  Lorentz structures \cite{Polyakov:2019lbq}: 
\bea
t_{\mu\nu}&=& 
\langle p', \sigma'| \hat{T}_{\mu\nu}(0)| p,\sigma \rangle = 
2P_\mu P_\nu \left[-\epsilon'\cdot\epsilon
  A_{0}(q^2)+\dfrac{P\cdot\epsilon'P\cdot\epsilon}{M^2}A_1(q^2)\right]
\nonumber\\ 
&+& 2\Big[P_\mu\left(\epsilon'_{\nu}P\cdot \epsilon+\epsilon_{\nu}
  P\cdot
  \epsilon'\right)+P_\nu\left(\epsilon'_{\mu}P\cdot\epsilon+\epsilon_{\mu}
  P\cdot\epsilon'\right)\Big]J(q^2) + \dfrac{1}{2}\left(q_\mu
  q_\nu-g_{\mu\nu} q^2\right)\left[\epsilon'\cdot\epsilon
  D_0(q^2)+\dfrac{P\cdot\epsilon' P\cdot\epsilon}{M^2}D_1(q^2)\right]
\nonumber\\ 
&+&\biggl[\dfrac{1}{2}q^2\left(\epsilon_{\mu}\epsilon'_{\nu}+\epsilon'_{\mu}\epsilon_{\nu}\right)-\left(\epsilon'_{\nu}q_\mu+\epsilon'_{\mu}q_\nu\right)\epsilon\cdot
P+\left(\epsilon_{\nu}q_\mu+\epsilon_{\mu}q_\nu\right)\epsilon'\cdot
P-4g_{\mu\nu}P\cdot\epsilon'P\cdot\epsilon\biggr] E(q^2)  
\,,
\label{EMTdef}
\eea
where $(p,\sigma)$ and $(p',\sigma')$ are the (four-momentum, spin
index) quantum numbers of the initial and final states, respectively,
$q=p'-p$ is the momentum transfer, $P=(p+p')/2$, while $
\epsilon'^\beta \equiv \epsilon^{\star\beta}(p',\sigma')$ and $
\epsilon^\beta \equiv \epsilon^\beta(p,\sigma) $ are the polarization
vectors.  Further, $M$ is a mass scale introduced to render the GFFs
dimensionless. It is often taken equal to the mass of the system
$m$. However, to avoid mixing of orders in the non-relativistic 
$1/m$-expansion,  we prefer
to distinguish between two mass parameters until the non-relativistic reduction is performed.  
  The one-particle states $|p,\sigma\rangle$ 
  satisfy the normalization condition
\begin{equation}
\langle  p',\sigma'| p ,\sigma\rangle=2\, p^0 \,(2\,\pi)^3\,\delta^3({\bf p}-{\bf p}')\delta_{\sigma\sigma'}.
\label{normalization}
\end{equation}

Up-to-and-including zeroth order in the $1/m$ expansion (corresponding to static approximation) for the kinematics with
$P^i=0$, we obtain the following expressions \cite{Panteleeva:2023evj}:  
\bea
t^{00}&=&2m^2\left[\delta_{\sigma'\sigma}\left(\mathcal{E}_0(-{\bf q}^2)+\dfrac{{\bf q}^2}{3M^2}\mathcal{E}_2(-{\bf q}^2)\right)-\dfrac{q_{\sigma'}q_\sigma}{M^2}\mathcal{E}_2(-{\bf q}^2)
\right] \,,\nonumber\\
t^{0i}&=&-m\mathcal{J}(-{\bf q}^2)\left(\delta_{i\sigma'}q_\sigma-\delta_{i\sigma}q_{\sigma'}\right)\,,\nonumber\\
t^{ij}&=&2\Bigg[\mathcal{D}_2(-{\bf q}^2)\Big(\delta_{ij}q_\sigma q_{\sigma'}-\dfrac{1}{2}q^i(q_\sigma\delta_{j\sigma'}+q_{\sigma'}\delta_{j\sigma})-\dfrac{1}{2}q^j(q_\sigma\delta_{i\sigma'}+q_{\sigma'}\delta_{i\sigma})+\dfrac{{\bf q}^2}{2}\left(\delta_{i\sigma}\delta_{j\sigma'}+\delta_{i\sigma'}\delta_{j\sigma}\right)\Big)\ \nonumber\\
&+&\left({\bf q}^2\delta_{ij}-q_iq_j\right)\left\{\delta_{\sigma'\sigma}\left(\mathcal{D}_0(-{\bf q}^2)-\dfrac{2}{3}\mathcal{D}_2(-{\bf q}^2)+\dfrac{{\bf q}^2}{3M^2}\mathcal{D}_3(-{\bf q}^2)\right)-\dfrac{q_{\sigma'}q_\sigma}{M^2}\mathcal{D}_3(-{\bf q}^2)
\right\}
\Bigg] \,,
\label{EMT_ME_Spin1}
\eea
where the combinations of the form factors are given by
\bea
\mathcal{E}_0(-{\bf q}^2)&=&A_0(-{\bf q}^2)-\dfrac{{\bf q}^2}{12 M^2}A_1(-{\bf q}^2) 
\,,\nonumber\\
\mathcal{E}_2(-{\bf q}^2)&=&\dfrac{A_1(-{\bf q}^2)}{4} 
\,,\nonumber\\
\mathcal{J}(-{\bf q}^2)&=& J(-{\bf q}^2) 
\,,\nonumber\\
\mathcal{D}_0(-{\bf q}^2)&=&\dfrac{D_0(-{\bf q}^2)}{4}+\dfrac{{\bf q}^2}{48  M^2} D_1(-{\bf q}^2)-\dfrac{E(-{\bf q}^2)}{3} 
\,,\nonumber\\
\mathcal{D}_2(-{\bf q}^2)&=&-\dfrac{E(-{\bf q}^2)}{2} 
\,,\nonumber\\
\mathcal{D}_3(-{\bf q}^2)&=&-\dfrac{D_1(-{\bf q}^2)}{16}
\,.
\eea
Above, we have chosen the Cartesian basis for three-dimensional polarization vectors.

\medskip

\begin{figure}
\epsfig{file=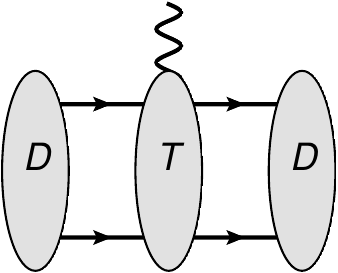, width=6cm}
\caption[]{\label{vertex} 
{Three-point function of the EMT operator and two interpolating fields of the deuteron. Ellipses with $D$ represent the amplitudes of the deuteron interpolating field interacting with a pair of nucleon fields, while
$T$ stands for the two-nucleon-irreducible part of the vertex function.  
Solid and  wavy lines denote the nucleons and the EMT insertion, respectively.}}
\end{figure}

Using Eq.~(\ref{EMT_ME_Spin1}), we extract below  the GFFs of the deuteron from the three-point function of the EMT operator $\hat T^{\mu\nu}$ and two interpolating fields of the deuteron,
\begin{equation}
G_{\sigma' \sigma}^{\mu\nu}(p',  p)=\int d^4 x d^4 y\, e^{-i p'\cdot
  y}  e^{i p \cdot x}\langle 0|T\left[{\cal D}_{\sigma}^\dagger(x) \hat T^{\mu\nu}(0)\,{\cal D}_{\sigma'}(y)\right]|0\rangle ,
\label{VertexD}
\end{equation}
using the LSZ reduction formula \cite{Kaplan:1998sz}
\begin{eqnarray}
\langle p',\sigma'|\hat T^{\mu\nu}| p,\sigma\rangle &=& 
-\frac{1}{Z}\,\left[\left(p^2-M_d^2\right)\left(p'^2-M_d^2\right)
                                                        G_{\sigma'\sigma}^{\mu\nu}(p',p)\right]_{p^2,
                                                        \, p'^2\to M_d^2},
\label{LSZ}
\end{eqnarray}
where $M_d=2m_N-E_b$ is the deuteron mass,  $E_b$ is its binding
energy, $m_N$ refers to the nucleon mass, while 
$Z$ is the residue of the deuteron propagator. 
The deuteron interpolating field in Eq.~(\ref{VertexD}) is defined as \cite{Kaplan:1998sz} 
\begin{equation}
{\cal D}_i\equiv N^T \mathcal{P}_i N=\sum_{\alpha,\beta,a,b=1}^2 N_{\alpha,a} \mathcal{P}_{i,ab}^{\alpha\beta} N_{\beta,b}
, \ \ \ \mathcal{P}_i\equiv \frac{1}{\sqrt{8}}\,{\sigma_2\sigma_i \tau_2},
\label{interfield}
\end{equation}
where $\alpha, \,\beta$ and $a, \,b$ are the spin and isospin indices, respectively.
Notice that observable quantities do not depend on a particular form of interpolating fields.
The two-point function of the deuteron interpolating fields is given by 
\begin{equation}
G_{\cal D}(p) \delta_{\sigma'\sigma}=\int d^4 x e^{-i p x} \langle 0|T\left[{\cal D}_{\sigma'}^\dagger(x) {\cal D}_\sigma(0)\right]|0\rangle = \delta_{\sigma'\sigma}\,\frac{i\, 2 M_d \, Z}{p^2-M_d^2+i\,\epsilon} + {\rm N.P.}\,,
\label{propagatorD}
\end{equation}
where ''N. P.'' stands for contributions without the
deuteron pole. 
The vertex function $G_{\sigma'\sigma}^{\mu\nu}(p',p)$ can be represented diagrammatically as shown in Fig.~\ref{vertex}. 

\section{The deuteron equation}
\label{sec3}

Below we briefly summarize the derivation of the equation for the
deuteron structure functions of Ref.~\cite{Epelbaum:2013naa}, applied
to the non-relativistic case.  
In non-relativistic low-energy EFT, the NN scattering amplitude is
obtained by solving the integral equation 
\begin{equation}
{T \left(
{\bf p'},{\bf p}\right)}{=}{ V \left(
{\bf p'},{\bf p}\right) - m_N \int \frac{d^3 {\bf k}}{(2\,\pi)^3}} {\frac{V \left(
{\bf p'},{\bf k}\right) \, T \left(
{\bf k},{\bf p}\right)}{ m_N E- {\bf k}^2+i\,\epsilon },}
\label{kad}
\end{equation}
where $E={\bf p}^2/m_N$   denotes the energy of two incoming nucleons in the
center-of-mass frame.

For our calculations, we need the amplitude of the deuteron
interpolating field interacting with a pair of nucleon fields. This
quantity in the rest frame of the deuteron is given by 
\begin{equation}
D_j({\bf p'}) = \mathcal{P}_j + m_N \int \frac{d^3{\bf k}}{(2\,\pi)^3} \,
\frac{\mathcal{P}_j \, T({\bf p'},{\bf k})}{
m_N E- {\bf k}^2+i\,\epsilon}\,,
\label{TD}
\end{equation}
where the NN scattering amplitude $T \left(
{\bf p'},{\bf p}\right)$ is obtained by solving Eq.~(\ref{kad}).
The amplitude $D \left( {\bf p'}\right)$ can be parameterized in terms of two
structure functions $\Delta_1$ and $\Delta_2$ via
\begin{equation}
D_{j}\left( {\bf p'}\right) = \Delta_1({\bf p'}^2) \mathcal{P}_j + p'_a p'_b\, \Delta_2({\bf p'}^2)\, \left(\sigma^a\right)^T \mathcal{P}_j \sigma^b  ,
\label{Dpar}
\end{equation}
where we do not show the isospin indices and terms resulting from anti-symmetrization. Notice that the structure functions $\Delta_1$ and $\Delta_2$
can be easily related to the $S$- and $D$-state components of the deuteron wave function (see also Ref.~\cite{Fachruddin:2001sb}).
To obtain equations for the structure functions $\Delta_{1,2}$ we parameterize the NN potential by
indicating explicitly the spin indices: 
\begin{eqnarray}
V_{\alpha\beta,\, \gamma\delta}\left({\bf p'} ,{\bf p}\right) & = & v^0
\left( {\bf p'} ,{\bf p} \right)
\delta_{\alpha\gamma}\delta_{\beta\delta}+v^1_a \left( {\bf p'} ,{\bf p}
\right) \,\left(\sigma^a_{\alpha\gamma}\delta_{\beta\delta} +
\delta_{\alpha\gamma}\sigma^a_{\beta\delta}\right) +v^2_{ab} \left( {\bf
p'} ,{\bf p}\right)
\,\sigma^a_{\alpha\gamma}\sigma^b_{\beta\delta}\,, 
\label{tpar}
\end{eqnarray}
where
\begin{eqnarray}
v^0({\bf p'},{\bf p}) & = & \nu_1({\bf p'},{\bf p})\,,\nonumber\\
v^1_a({\bf p'},{\bf p}) & = & i\,\epsilon^{abc} p^b p'^c
\nu_3({\bf p'},{\bf p})\,, \\
v^2_{ab}({\bf p'},{\bf p}) & = & \delta_{ab}\, \nu_2({\bf p'},{\bf p}) +p'^a p'^b\,
\nu_5({\bf p'},{\bf p})+ p^a
p^b\, \nu_6({\bf p'},{\bf p}) +  (p^a
p'^b+ p'^a p^b)\,\nu_4({\bf p'},{\bf p}),  \nonumber
\label{fversusampl}
\end{eqnarray}
and $\nu_i({\bf p'},{\bf p})$ are scalar functions of ${\bf
p'}^2$, ${\bf p}^2$ and ${\bf p'}\cdot{\bf p}$. 
For the structure functions, we obtain the following system of integral equations:
\begin{eqnarray}
\Delta_1({\bf p}^2) & = & 1+ m_N \int\frac{d^3{\bf k}}{(2\,\pi)^3}\, G({\bf k})
\biggl\{  \Delta_1({\bf k}^2) \left[\nu_1({\bf p},{\bf k})+\nu_2({\bf p},{\bf k})+C_1 \nu_6({\bf p},{\bf k})\right]    \nonumber  \\
&+& \Delta_2({\bf k}^2) \biggl[C_1\left(\nu_1({\bf p},{\bf k})+\nu_2({\bf p},{\bf k})\right)+2 ({\bf p}\cdot {\bf k}) \nu_3({\bf p},{\bf k})+2\, {\bf k}^2 ({\bf p}\cdot {\bf k})\, \nu_4({\bf p},{\bf k})   \nonumber       \\
&+&  \left[({\bf p}\cdot {\bf k})^2-C_1 {\bf p}^2\right]\nu_5({\bf p},{\bf k})+({\bf k}^2)^2 \nu_6({\bf p},{\bf k})\biggr]\biggr\},    \nonumber  \\
\Delta_2({\bf p}^2) & = &  m_N \int \frac{d^3{\bf k}}{(2\,\pi)^3}\, G({\bf k})\,
\biggl\{\Delta_1({\bf k}^2) [2 B\, \nu_4({\bf p},{\bf k})+C_2 \nu_6({\bf p},{\bf k})+\nu_5({\bf p},{\bf k})]    \nonumber       \\
&+& \Delta_2({\bf k}^2)
\biggl[C_1 \nu_5({\bf p},{\bf k})
-2 B {\bf k}^2 \nu_3({\bf p},{\bf k})  + C_2 \left(\nu_1({\bf p},{\bf k})
+ \nu_2({\bf p},{\bf k})+2\, ({\bf p}\cdot {\bf k})\, \nu_3({\bf p},{\bf k})\right)
\biggr]\biggr\},
\label{dsystem2}
\end{eqnarray}
where 
\begin{equation}
B  \equiv  \frac{{\bf p}\cdot {\bf k}}{{\bf p}^2},\quad
C_1 \equiv  \frac{1}{2}\,\left[{\bf k}^2- \frac{({\bf p}\cdot {\bf k})^2}{{\bf p}^2}\right],\quad
C_2  \equiv  \frac{3\,({\bf p}\cdot {\bf k})^2-{\bf k}^2 {\bf p}^2}{2\,{\bf p}^4}.
\label{BCDef}
\end{equation}

\medskip

For our calculations of the deuteron GFFs we consider the regulated leading-order NN potential of chiral EFT  given by
\begin{equation}
V_0\left(
{\bf p'},{\bf p}\right) = \left(C_S+C_T\, {\bf\sigma_1}\cdot{\bf\sigma_2} \right) 
\frac{\Lambda^4}{({\bf p'}^2+\Lambda^2)({\bf p}^2+\Lambda^2)}
-\frac{g_A^2}{4\,F_\pi^2}\ {\bf \tau_1}\cdot{\bf\tau_2} \
\frac{{\bf\sigma_1}\cdot ({\bf p}'-{\bf p})\,{\bf\sigma_2}\cdot ({\bf p}'-{\bf p})}{({\bf p}'-{\bf p})^2+M_\pi^2} \frac{\Lambda^2-M_\pi^2}{({\bf p}'-{\bf p})^2+\Lambda^2} \,,
\label{LOV}
\end{equation}
which corresponds to
\begin{eqnarray}
\nu_1({\bf p'},{\bf p}) & = & C_S \, \frac{\Lambda^4}{({\bf p'}^2+\Lambda^2)({\bf p}^2+\Lambda^2)} \,, \nonumber \\
\nu_2({\bf p'},{\bf p}) & = & C_T \, \frac{\Lambda^4}{({\bf p'}^2+\Lambda^2)({\bf p}^2+\Lambda^2)} \,, \nonumber \\
\nu_3({\bf p'},{\bf p}) & = & 0\,, \nonumber \\
\nu_4({\bf p'},{\bf p}) & = & \frac{g_A^2}{4\,F_\pi^2}\ {\bf \tau_1}\cdot{\bf\tau_2} \
\frac{1}{({\bf p}'-{\bf p})^2+M_\pi^2} \frac{\Lambda^2-M_\pi^2}{({\bf p}'-{\bf p})^2+\Lambda^2} \,, \nonumber \\
\nu_5({\bf p'},{\bf p}) & = & -\frac{g_A^2}{4\,F_\pi^2}\ {\bf \tau_1}\cdot{\bf\tau_2} \
\frac{1}{({\bf p}'-{\bf p})^2+M_\pi^2} \frac{\Lambda^2-M_\pi^2}{({\bf p}'-{\bf p})^2+\Lambda^2} \,, \nonumber \\
\nu_6({\bf p'},{\bf p}) & = & -\frac{g_A^2}{4\,F_\pi^2}\ {\bf \tau_1}\cdot{\bf\tau_2} \
\frac{1}{({\bf p}'-{\bf p})^2+M_\pi^2} \frac{\Lambda^2-M_\pi^2}{({\bf p}'-{\bf p})^2+\Lambda^2} \,, \label{NLOCVpar}
\end{eqnarray}
where $C_S$ and $C_T$ are low-energy coupling constants of the
leading-order NN contact interaction Lagrangian
\cite{Weinberg:rz,Weinberg:um}, $g_A$ is the nucleon axial charge, while
$M_\pi$ and $F_\pi$ are the mass and the decay constant of the pion,
respectively. 
In the above expressions, we have introduced a smooth cutoff with a
parameter $\Lambda$ to regularize ultraviolet divergences. 
The employed regulator for the one-pion-exchange potential does
not modify its longest-range part (i.e., the left-hand cut in the
on-shell partial-wave two-nucleon amplitude closest to the physical region) and the physics related 
to it~\cite{Reinert:2017usi,Gasparyan:2012km}.
For the same reason, we adopted the effective value of the axial coupling constant $g_A=1.29$~\cite{Epelbaum:2004fk}
that takes into account the Goldberger-Treiman discrepancy.
Note that our regularization procedure violates EMT conservation,
which could potentially lead to the appearance of 
contributions to the form factors proportional to positive powers of
$\Lambda$, which cannot be absorbed
by counter terms of the effective Lagrangian allowed by symmetries.
However, as we will see from our results, such effects appear to be rather small.
A more systematic approach can be developed by applying a symmetry-preserving regularization  \cite{Djukanovic:2004px,Djukanovic:2006mc,Krebs:2023gge}, which however goes beyond 
the scope of the current work. 
We set the numerical value of the cutoff parameter of the order of the
EFT expansion breakdown scale $\Lambda \sim \Lambda_b\sim 400-600$~MeV~\cite{Epelbaum:2014efa,Furnstahl:2015rha,Epelbaum:2019zqc,Epelbaum:2019wvf}.
To analyze the sensitivity of our results to the choice of the cutoff, we vary it within this range $\Lambda\in(400,600)$~MeV.
For the central cutoff value, we choose $\Lambda=500$~MeV.
	
\medskip

To calculate the residue of the two-point function of the deuteron interpolating fields it is useful to write 
the dressed deuteron propagator $G_{\cal D}$  using Eq.~(\ref{Dpar}) as
\begin{equation}
G_{\cal D}\left(E,{\bf 0}\right)= m_N \int \frac{d^3 {\bf
    k}}{(2\,\pi)^3} \frac{\tilde \Delta_1 \left({\bf k}^2\right)}{
m_N E- {\bf k}^2 +i\,\epsilon}+{\rm N.\, P.},
\label{2pointf}
\end{equation}
where $ \tilde \Delta_1 \left({\bf k}^2\right) = \Delta_1 \left({\bf k}^2\right) +\frac{{\bf k}^2}{3}\, \Delta_2 \left({\bf k}^2\right)$ and "N.\, P." stands for the non-pole part. 

\begin{figure}
\epsfig{file=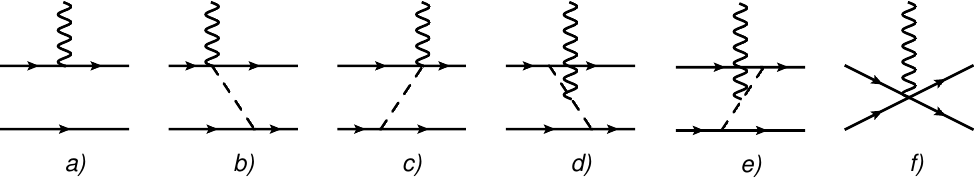, width=14cm}
\caption[]{\label{vertexTree} 
Tree-level diagrams contributing to the vertex function of the EMT operator. Diagrams where the graviton couples to the second nucleon line are not shown. Solid, dashed and wavy lines correspond to nucleons, pions and gravitons, respectively.}
\end{figure}

\section{Calculation of the form factors}
\label{calc}

In this section we calculate the matrix element of the EMT  in the
Breit frame for the initial and final deuteron states with the quantum
numbers $(-{\bf q}/2 ,\sigma)$ and $({\bf
  q}/2 ,\sigma')$.
For the matrix element, we obtain
\bea
\label{Int1}
\langle {\bf q}/2 ,\sigma' |\hat T_{\mu\nu}|-{\bf q}/2 ,\sigma  \rangle=m_N^2 \! \!
\int \! \dfrac{d^3 k_1}{(2\pi)^3} \dfrac{d^3 k_2}{(2\pi)^3} \dfrac{ D_{\sigma',cd}^{\gamma\delta}({\bf k}_2) \, T^{cd,\gamma\delta;ab,\alpha\beta}_{\mu\nu}
\left(\frac{\bf q}{4}+ {\bf k}_2, \frac{\bf q}{4}-{\bf k}_2;{\bf k}_1- \frac{\bf q}{4},-{\bf k}_1- \frac {\bf q}{4} \right)
D^{\dagger,\alpha\beta}_{\sigma,ab}({\bf k}_1) 
}{ ({\bf k}_1^2+p_B^2)({\bf k}_2^2+p_B^2) } ,
\eea
where $p_B= \sqrt{m_N E_b}$ is the deuteron binding momentum and the
arguments of the amplitude in the integrand correspond to the
individual momenta of both nucleons. Comparing the results of
calculating the expression of Eq.~(\ref{Int1}) with the
parameterization of Eq.~(\ref{EMT_ME_Spin1}) we extract the GFFs of
the deuteron in the static approximation. 

\medskip

To calculate the order-by-order approximations to the deuteron matrix
element of the EMT we apply the standard Weinberg power counting for the few-body sector of chiral EFT \cite{Weinberg:rz,Weinberg:um} to the integral in Eq.~(\ref{Int1}), see, e.g., Refs.~\cite{Epelbaum:2007us, Krebs:2020pii} for details.
According to this power counting, the pion mass and external three-momenta (divided by $\Lambda_b$)
count as of order one, each internal pion (nucleon) line counts as of order minus two (minus one),
the nucleon mass counts as of order minus one (NN power counting, see
Refs.~\cite{Weinberg:rz,Weinberg:um,Epelbaum:2008ga,Epelbaum:2019kcf} for details)
and a noninteracting spectator nucleon counts as of order minus three.
Each pionic loop adds four to the overall chiral order.
Interaction vertices originating from the effective Lagrangian of order $N$ count as of chiral order $N$, 
while the vertices corresponding to EMT have different orders for different components.
The EMT corresponding to the first-order Lorentz-invariant Lagrangian
generates the vertices with contributions starting with the order minus one, all other $N$-th order Lagrangians lead to $N$-th or higher order
contributions in the corresponding EMT vertices, modulo (enhanced)
factors of the nucleon mass in terms with derivatives acting on
nucelon field.
The relevant single-nucleon effective Lagrangian and the corresponding expression for the energy-momentum tensor can be found in Ref.~\cite{Alharazin:2020yjv}.

In our calculation, we include all diagrams shown in Fig.~\ref{vertexTree},
which contain contributions to the vertex function up to zeroth order
in all components of the $T_{\mu\nu}$ part in the integrand of
Eq.~(\ref{Int1}).  
Note that chiral expansion for different components of the deuteron matrix element of EMT
starts at different orders.
In particular, the expansion for the ${00}$-th component starts at order $-4$, whereas the leading contributions
to the ${0i}$-th and  ${ij}$-th components come from orders $-2$ and $0$, respectively.

In a fully consistent approach, one would have to include also higher-order corrections to
the nucleon-nucleon potential ($T$-matrix) as well as relativistic corrections to the nucleon propagator, at least for $T_{00}$ and $T_{0i}$.
Obviously, in the present exploratory study, such an unnecessary complication would
provide us with no additional information. Therefore, we stick to the leading-order
NN potential and drop the relativistic corrections to the nucleon propagator for all gravitational form factors.

\medskip

The leading contribution to the form factors
is given by diagram a) in Fig.~\ref{vertexTree} (and its partner
diagram where the graviton couples to another nucleon) with one EMT insertion in a single nucleon line. It has the following form:
\be\label{Int2}
\langle {\bf q}/2 ,\sigma' |\hat T^{\rm LO}_{\mu\nu}|-{\bf q}/2 ,\sigma  \rangle =  4 m_N^2\int\dfrac{d^3 k}{(2\pi)^3}
\dfrac{D_{\sigma'}({\bf k}+{\bf q}/4) T_{\mu\nu,a}({{\bf k}+{\bf q}/2,{\bf k}-{\bf q}/2} )D^\dagger_\sigma({\bf k}-{\bf q}/4)}
{\left[({\bf k}+{\bf q}/4)^2+p_B^2\right]\left[({\bf k}-{\bf q}/4)^2+p_B^2\right]} \,,
\ee 
where, up to the accuracy of our calculation, we have: 
\bea
T_{00,a} ({\bf k} + {\bf q} /2, {\bf k} - {\bf q} /2) &=&m_N + \frac{{\bf k}^2}{2m_N}  - \frac{i \epsilon^{lmn} \sigma^l k^m q^n}{4m_N} 
+c_8 \,\frac{{\bf q}^2}{4} +2 c_9 \,{\bf q}^2
\,, \nonumber\\
T_{0i,a} ({\bf k} + {\bf q} /2, {\bf k} - {\bf q} /2) &=& k_i +
\frac{i \epsilon_{ilm} \sigma^l q^m}{4} 
+ \frac{c_9}{m_N} \left[ {\bf q}^2 k_i +({\bf k\cdot q}) q_i\right]
\,, \nonumber\\
T_{ij,a} ({\bf k} + {\bf q} /2, {\bf k} - {\bf q} /2) &=& \frac{k_i k_j}{m_N}  + \frac{i  \sigma^l q^m}{4m_N}  \left( k_i \epsilon_{jlm} + k_j \epsilon_{ilm} \right) 
- \frac{c_8}{4}\left[ {\bf q}^2 \delta_{ij} - q_i q_j \right]
\,. 
\label{diaga}
\eea
Here, $c_8$ and $c_9$ are coupling constants of the subleading
pion-nucleon Lagrangian in curved spacetime \cite{Alharazin:2020yjv}
and ${\bf k} - {\bf q} /2$ and ${\bf k} + {\bf q} /2$ denote
three-momenta of the incoming and outgoing states of the nucleon,
respectively. Currently, there is no way to estimate the value of the
low-energy constant $c_9$, so we set $c_9=0$
in all numerical calculations. In the power counting scheme we employ, the disconnected
contributions 
stemming from the diagram a) of Fig.~\ref{vertexTree} 
are of
orders $-4$, $-2$, $-1$ and $0$.

\begin{figure}[tb]
\begin{minipage}[h]{0.495\linewidth}
\center{\includegraphics[width=0.8\linewidth]{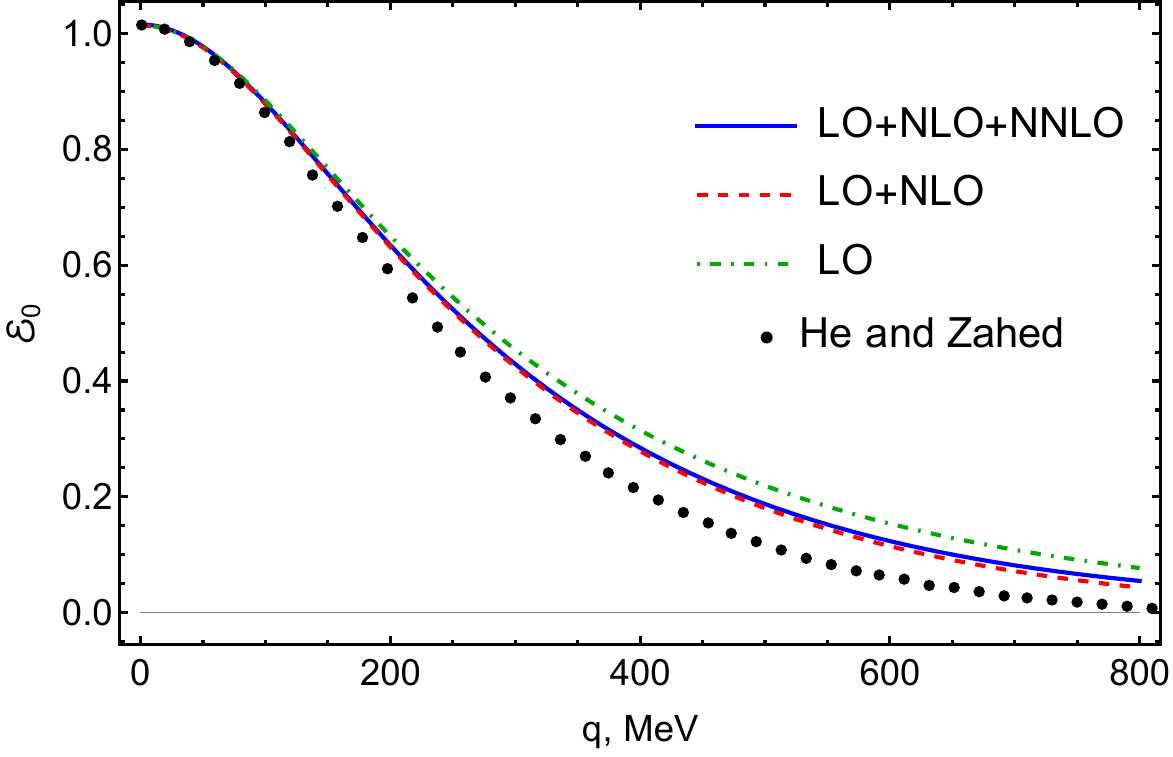}} \\
\end{minipage}
\hfill
\begin{minipage}[h]{0.495\linewidth}
\center{\includegraphics[width=0.8\linewidth]{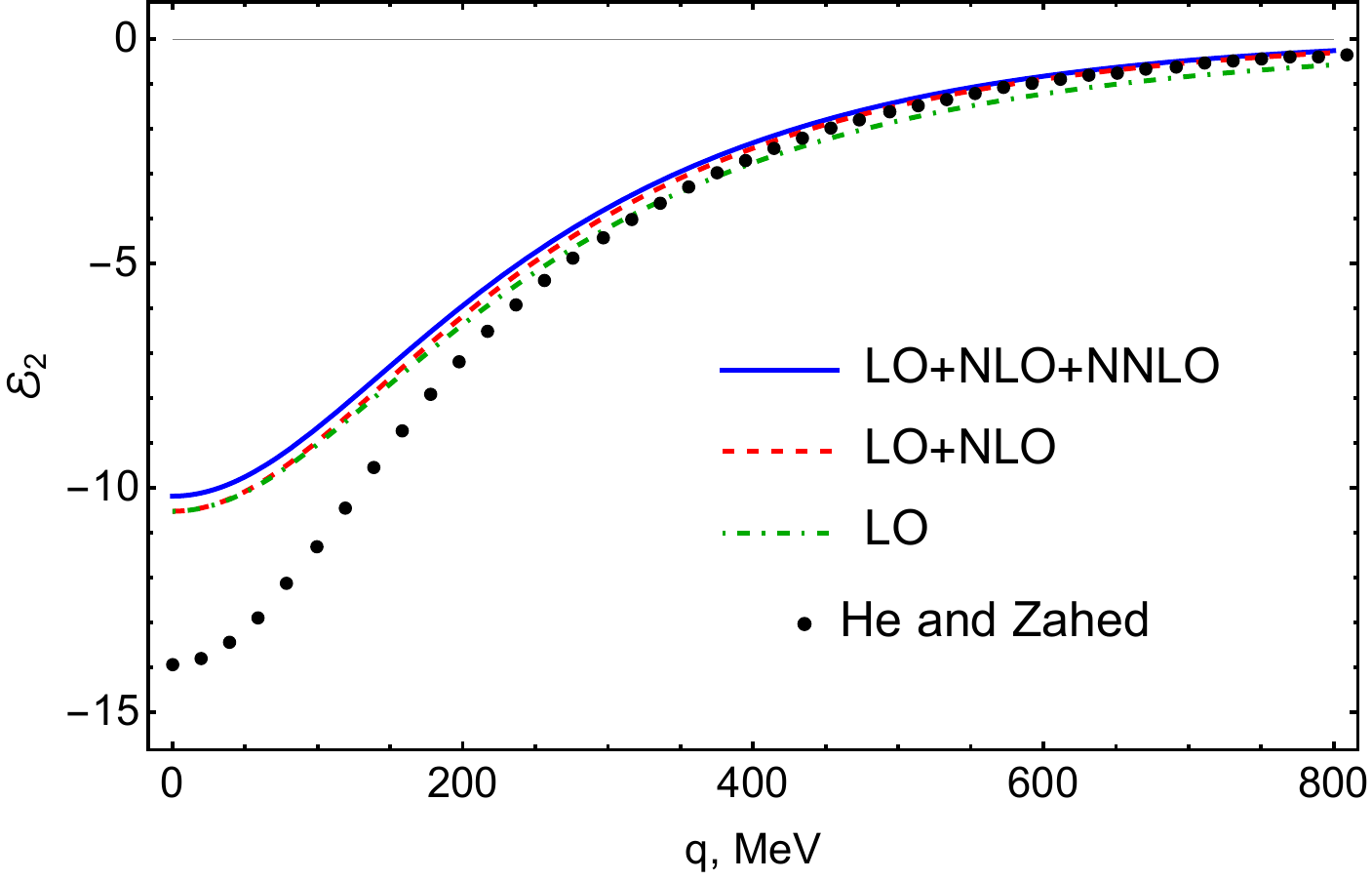}} \\
\end{minipage}
\vfill
\center{\includegraphics[width=0.4\linewidth]{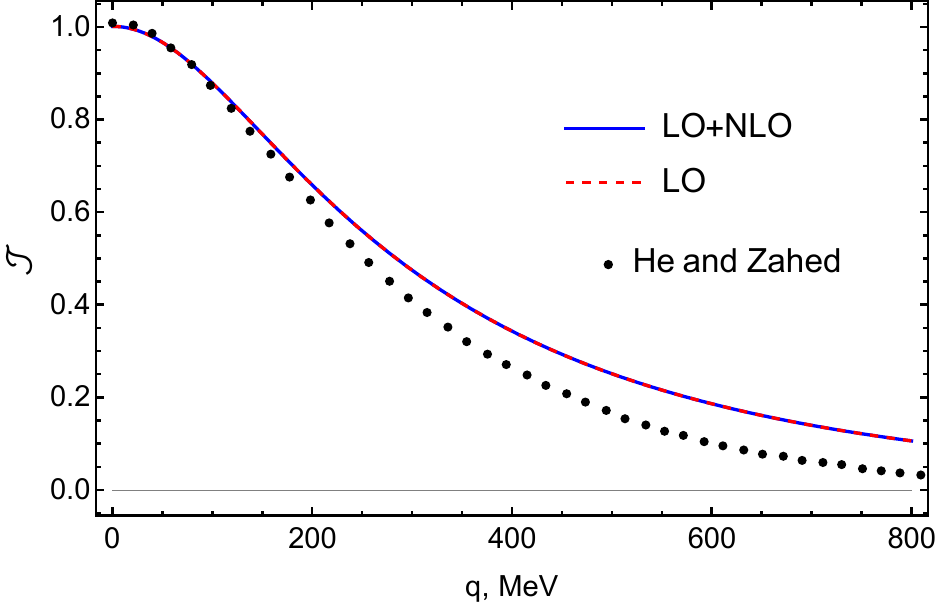}} \\
\caption{Numerical results for the gravitational form factors ${\cal E}$ and ${\cal J}$
of the deuteron using the cutoff parameter $\Lambda=500~ \text{MeV}$.  
Definitions of the EFT orders are given in text. 
We compare our results with the GFFs by He and Zahed given in Fig.~3 of Ref.~\cite{He:2024vzz}.  Notice that our two curves for ${\cal J}$ coincide due to vanishing NLO contribution to this form factor.}
\label{fig1ComPions}
\end{figure}

 The connected two-nucleon diagrams in
Fig.~\ref{vertexTree} are all of chiral order zero.
The regularized contributions of diagrams b) and c) to the vertex function 
$T_{\mu\nu}( {\bf k}_2+{\bf q}/4, -{\bf k}_2+{\bf q}/4;{\bf k}_1-{\bf
  q}/4,-{\bf k}_1- {\bf q}/4)$ at the accuracy of our calculations have
the following form (graviton coupling to both, the upper and lower vertices):
\bea
T_{00,b+c} & = & - \frac{ g_A^2 \, {\bf \tau_1}\cdot{\bf\tau_2}\,}{F_\pi^2} 
\frac{   \sigma_1\cdot \tilde {\bf k} \, \sigma_2\cdot \tilde {\bf k}}{\tilde {\bf k}^2+M_\pi^2} \frac{\Lambda^2-M_\pi^2}{ \tilde {\bf k}^2+\Lambda^2 } 
\,,\nonumber\\
T_{0i,b+c} & = & 0  \,,\\
T_{ij,b+c} & = &  - \frac{ g_A^2 \, {\bf \tau_1}\cdot{\bf\tau_2}\,}{4 F_\pi^2} 
\Biggl[ 
\frac{  \sigma_1\cdot \tilde {\bf k} }{\tilde {\bf k}^2+M_\pi^2} \, (-2\delta_{ij} \, \sigma_2\cdot \tilde {\bf k} + \tilde k^i \sigma_2^j + \tilde k^j \sigma_2^i ) 
+
\frac{ \sigma_2\cdot \tilde {\bf k} }{\tilde {\bf k}^2+M_\pi^2} \, (-2\delta_{ij} \, \sigma_1\cdot \tilde {\bf k} + \tilde k^i \sigma_1^j + \tilde k^j \sigma_1^i )
\Biggr]  \frac{\Lambda^2-M_\pi^2}{ \tilde {\bf k}^2+\Lambda^2 }  \,,
\nonumber
\label{HBres45}
\eea 
The regularized contributions of diagrams d) and e) have the form: 
\bea
T_{00,d+e} & = & \frac{ g_A^2 \, {\bf \tau_1}\cdot{\bf\tau_2}\,}{4 F_\pi^2} \frac{ \sigma_1\cdot \bar {\bf k} \, \sigma_2 \cdot \tilde {\bf k} }{(\bar {\bf k}^2+M_\pi^2)(\tilde {\bf k}^2+M_\pi^2)} 
\left[ M_\pi^2 + \bar {\bf k} \cdot \tilde {\bf k}  \right] \frac{\Lambda^2-M_\pi^2}{ \bar {\bf k}^2+\Lambda^2 } \frac{\Lambda^2-M_\pi^2}{ \tilde {\bf k}^2+\Lambda^2 }  \,,\nonumber\\
T_{0i,d+e} & = & 0  \,,\nonumber\\
T_{ij,d+e} & = & \frac{ g_A^2 \, {\bf \tau_1}\cdot{\bf\tau_2}\,}{4 F_\pi^2} \frac{ \sigma_1\cdot \bar {\bf k} \, \sigma_2 \cdot \tilde {\bf k}}{(\bar {\bf k}^2+M_\pi^2)(\tilde {\bf k}^2+M_\pi^2)} 
\left[ -\delta_{ij}(M_\pi^2+\bar {\bf k}\cdot \tilde {\bf k}) +\bar k^i \tilde k^j +\bar k^j \tilde k^i \right]  \frac{\Lambda^2-M_\pi^2}{ \bar {\bf k}^2+\Lambda^2 } \frac{\Lambda^2-M_\pi^2}{ \tilde {\bf k}^2+\Lambda^2 } \,,
\label{HBres3}
\eea
and the result of diagram f), which corresponds to the coupling of the EMT to 
the leading order NN contact interaction vertex, reads:
\bea
T_{\mu\nu,f} & = & -2 (C_S+C_T) g^{\mu\nu}\,,
\label{diagh}
\eea
where 
we used the shorthand notation with $\bar {\bf k}={\bf k}_1-{\bf
  k}_2-{\bf q}/2$ and $\tilde {\bf k}={\bf k}_1-{\bf k}_2+{\bf q}/2$.   
Notice that the one-loop corrections to the single-nucleon EMT are also formally of order zero.
However, their contributions result merely in the renormalization of
the nucleon mass, nucleon field and the $c_8$ and $c_9$  coupling constants.
We have not shown the one-pion-exchange two-nucleon irreducible diagrams where the EMT couples to a single nucleon line, because their contributions are canceled by $1/m_N$ corrections to the one-pion-exchange nucleon-nucleon potential. This is completely analogous to
the well-known cancellation of the analogous time-ordered perturbation theory contribution to the three-nucleon force 
at NLO by iterations of the one-pion-exchange potential including
the $1/m_N$ correction, see, e.g., Ref.~\cite{vanKolck:1994yi}.

\begin{figure}[tb]
\begin{minipage}[h]{0.495\linewidth}
\center{\includegraphics[width=0.8\linewidth]{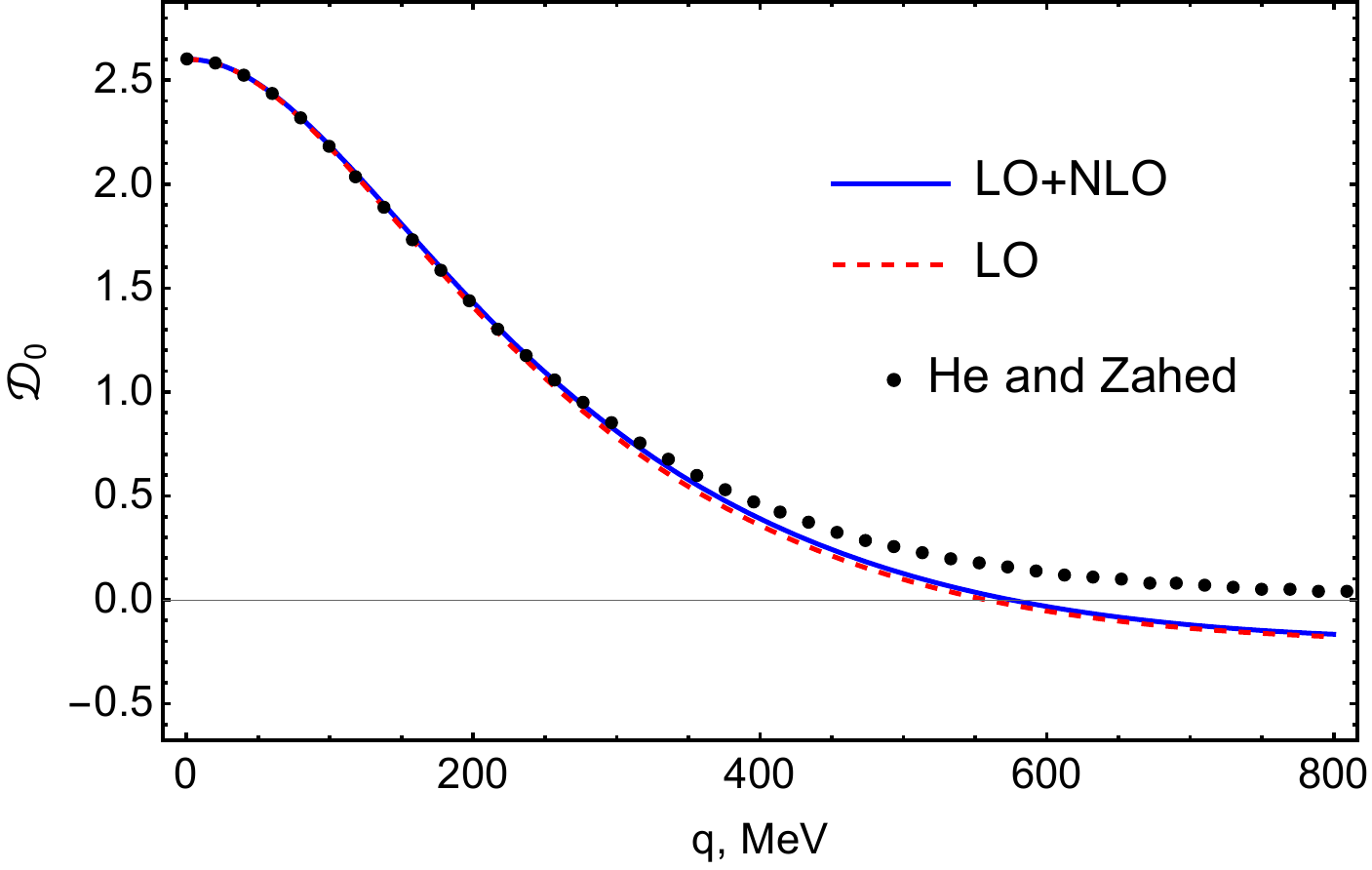}} \\
\end{minipage}
\hfill
\begin{minipage}[h]{0.495\linewidth}
\center{\includegraphics[width=0.8\linewidth]{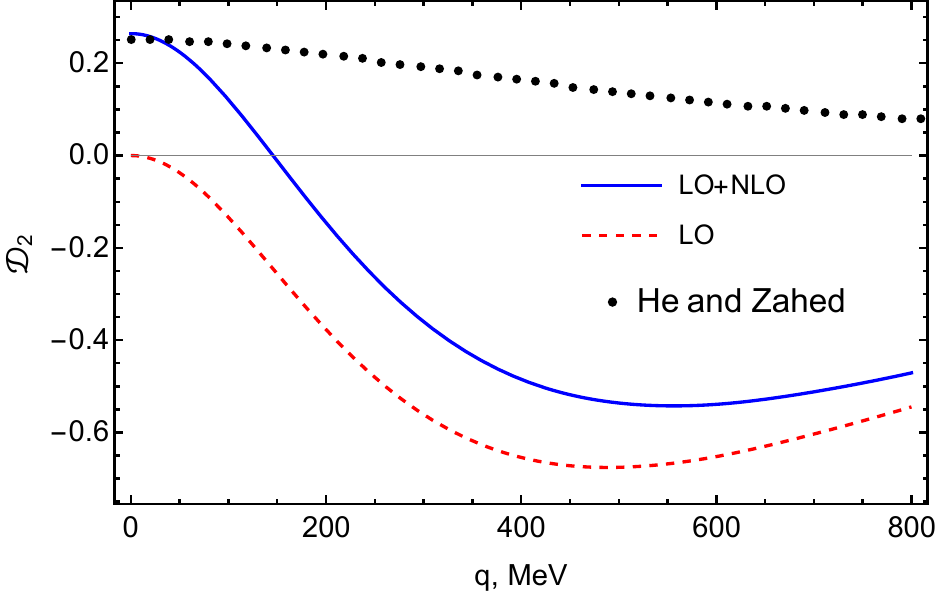}} \\
\end{minipage}
\vfill
\begin{minipage}[h]{0.495\linewidth}
\center{\includegraphics[width=0.8\linewidth]{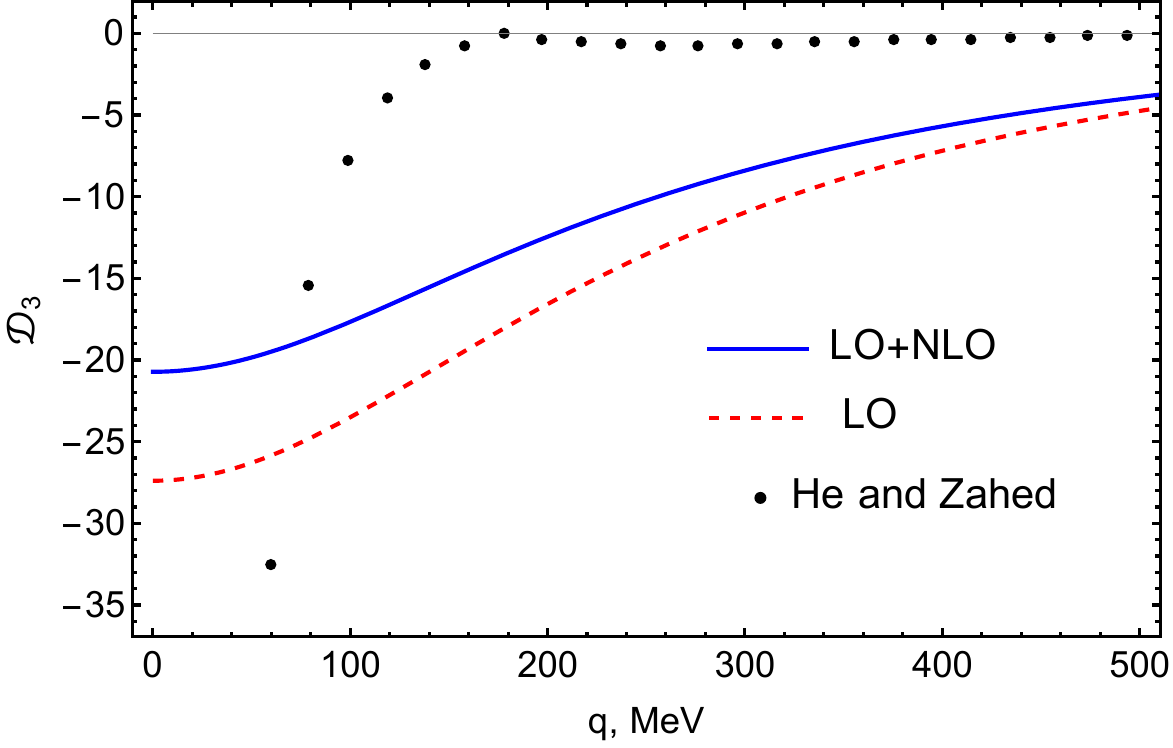}} \\
\end{minipage}
\hfill
\caption{Numerical results for the gravitational form factors ${\cal
    D}_i$ using the cutoff parameter $\Lambda=500~ \text{MeV}$.
Definitions of the EFT orders are given in text.
We compare our results with the GFFs by He and Zahed given in Fig.~3 of Ref.~\cite{He:2024vzz}. }
\label{fig7Com}
\end{figure}

\medskip

Using the above expressions, we calculated the GFFs of the deuteron by
solving the integral equations for the deuteron amplitudes and
calculating the matrix elements of EMT numerically.  
We fixed the coupling constant of the $S$-wave nucleon-nucleon contact
interaction by reproducing the binding energy of the deuteron. 
In Figs.~\ref{fig1ComPions} and \ref{fig7Com}, we show our results
using $\Lambda=500$ MeV together with the results of
Ref.~\cite{He:2024vzz} for the deuteron GFFs in the parameterization
of Eq.~(\ref{EMT_ME_Spin1}).\footnote{ 
Notice that while we plot the figures for $q$ up to $800 \, {\rm MeV}$
to compare to Ref.~\cite{He:2024vzz}, our chiral EFT results cannot be
trusted at such large values of the momentum transfer.} 
The notation for chiral orders of various form factors follows from the explicit expressions
for the matrix elements in Eqs.~\eqref{diaga}-\eqref{HBres45}:
\begin{itemize}
\item{For the form factors ${\cal E}_0$ and ${\cal E}_2$, the LO contribution involves diagram a)
	apart from the $c_8$ term that forms the NLO contribution, while the 
	remaining diagrams constitute the NNLO terms.}
\item{For the form factor ${\cal J}$, the whole contribution up to the accuracy of our calculation is generated
    by diagram a).}
\item{For the form factors ${\cal D}_0$, ${\cal D}_2$ and ${\cal D}_3$, 
	the LO term emerges from diagram a) proportional to $c_8$,
        while the remaining nonvanishing diagrams contribute at NLO.}
\end{itemize}

Our calculated form factors ${\cal E}_0$, ${\cal E}_2$ and ${\cal J}$
show, as functions of $q = | {\bf q}|$,  a similar behavior to those of Ref.~\cite{He:2024vzz}.
Note that the deviation from $1$ of the ${\cal E}_0$  form factor
at $q=0$ (corresponding to the mass of the deuteron) is tiny, which indicates that 
the effect of the  violation of the EMT conservation due to the noninvariant regularization and the neglected relativistic corrections is rather small. 
We fixed the value of the coupling constant $c_8$ to $c_8=-2.77\ {\rm GeV}^{-1}$, such that our calculated value of ${\cal D}_0(0)$ coincides with that of Ref.~\cite{He:2024vzz}. 
The extracted value of $c_8$ is of natural size, which allows us to analyze convergence of the chiral expansion.
The resulting $q$-dependence of our form factor ${\cal D}_0$ is very
similar to that of Ref.~\cite{He:2024vzz}. On the other hand, our
curve for ${\cal D}_2$ has a different shape, while the form factor ${\cal D}_3$
shows qualitatively different behavior. This is because the results of
Ref.~\cite{He:2024vzz}, when recalculated in the GFFs of the
parameterization of Eq.~(\ref{EMT_ME_Spin1}), lead to ${\cal D}_3$
which has a singular behavior at the origin, while this is not the
case for our results. We checked that our numerical results show a
very mild cutoff dependence for values of $\Lambda$ between $\sim 400$
and $600$ MeV. 
The variation of the form factors is comparable with or smaller than the highest-order contribution considered.
As argued above, residual cutoff dependence is expected to be further reduced if a symmetry-preserving regulator is used.

The curves in Figs.~\ref{fig1ComPions},~\ref{fig7Com} demonstrate a reasonable convergence rate of the chiral expansion for the deuteron gravitational
form factors when going from the LO to NLO and then to NNLO results.
The only exception of the rapid convergence is the small-$q$ region of the form factor ${\cal D}_3$,
which is caused by the fact that the leading-order term proportional to the $c_8$ coupling constant vanishes at $q=0$.
One of the factors that accelerate the convergence is a small typical
momentum of the nucleons inside the deuteron (due to its small binding energy), 
which enters the convolution integrals in Eqs.~\eqref{Int1} and \eqref{Int2}.

\section{Summary}
\label{sec4}

In this work, we have calculated the GFFs of the deuteron in the framework of chiral effective field theory with pions and nucleons as dynamical degrees of freedom. 
We extracted the GFFs by applying the standard LSZ formalism to the three-point function of the EMT operator and the deuteron interpolating fields. 
To obtain the deuteron "wave function" in momentum space, we modified
the system of integral equations for the deuteron of
Ref.~\cite{Epelbaum:2013naa} by adjusting it to our non-relativistic
approach. To regularize the nucleon-nucleon potential consisting of
the contact interaction and the one-pion-exchange parts, as well as
the two-nucleon diagrams contributing to the EMT, we applied a smooth cutoff
regularization.  

We solved the integral equations for the deuteron, calculated the
matrix element of the EMT numerically, and compared the obtained GFFs
of the deuteron to the results of Ref.~\cite{He:2024vzz} by
recalculating the latter in terms of our parameterization. We have
fixed the free parameter $c_8$ that appears in our calculations by fitting
it to the value of ${\cal D}_0(0)$ from Ref.~\cite{He:2024vzz} and determined
the coupling constant of the $S$-wave nucleon-nucleon contact
interaction by reproducing the binding energy of the deuteron. 
The results of two calculations are found to show a similar behavior as
functions of the momentum transfer for ${\cal E}_0$, ${\cal E}_2$,
${\cal D}_0$ and ${\cal J}$, while  
the GFF for ${\cal D}_2$ shows a different pattern. As for
the form factor ${\cal D}_3$, the (recalculated) result of
Ref.~\cite{He:2024vzz} has a singular behavior  
for vanishing $q$ while ${\cal D}_3(0)$ is finite in our case. 
We have observed a rather rapid convergence of the chiral expansion for the gravitational form factors of the deuteron
and a mild cutoff dependence of the results.

\acknowledgements

We thank the authors of Ref.~\cite{He:2024vzz} for providing us with
their calculated numerical results for the deuteron GFFs.
This work was supported in part 
by ERC  NuclearTheory (grant No. 885150), by BMBF (Grant No. 05P21PCFP1),
by the MKW NRW under the funding code NW21-024-A,
by
DFG and NSFC through funds provided to the Sino-German CRC 110
“Symmetries and the Emergence of Structure in QCD” (NSFC Grant
No. 11621131001, DFG Project-ID 196253076 - TRR 110), by the Georgian Shota Rustaveli National
Science Foundation (Grant No. FR-23-856)
and by the EU Horizon 2020 research and
innovation program (STRONG-2020, grant agreement No. 824093).

\end{document}